\documentclass[pdflatex,sn-mathphys-num]{sn-jnl}

\usepackage{geometry}
\usepackage{graphicx}%
\usepackage{multirow}%
\usepackage{amsmath,amssymb,amsfonts}%
\usepackage{amsthm}%
\usepackage{amsmath, amssymb}
\usepackage{mathrsfs}%
\usepackage[title]{appendix}%
\usepackage{xcolor}%
\usepackage{textcomp}%
\usepackage{manyfoot}%
\usepackage{siunitx}
\usepackage{booktabs}%
\usepackage{algorithm}%
\usepackage{algorithmicx}%
\usepackage{algpseudocode}%
\usepackage{listings}%

\theoremstyle{thmstyleone}%
\theoremstyle{thmstyletwo}%

\theoremstyle{thmstylethree}%

\begin{document}

\title[Article Title]{Advances in Prebiotic Chemistry: the potential of Analog Computing and Navier-Stokes Nernst-Planck (NPNS) Modeling in Organic Electronics Technologies (OECTs)}


\author[1]{\fnm{Giuseppe} \sur{De Giorgio}}

\author[1]{\fnm{Pasquale} \sur{D’Angelo}}\email{pasquale.dangelo@ cnr.it}

\author[1]{\fnm{Giuseppe} \sur{Tarabella}}\email{giuseppe.tarabella@cnr.it}

\author[2]{\fnm{Saksham} \sur{Sharma}}

\affil[1]{\orgdiv{Institute of materials for Electronics and Magnetism (IMEM-CNR)}, \orgaddress{\street{Parco Area delle Scienze 37A}, \city{Parma}, \postcode{43124}, \state{Emilia-Romagna}, \country{Italy}}}

\affil[2]{\orgdiv{Cambridge Centre for Physical Biology (CCPB)}, \orgaddress{\street{Wilberforce Road}, \city{Cambridge}, \postcode{CB3 0WA}, \state{Cambridgeshire}, \country{United Kingdom}}}


\abstract{In this article, we attempt to make a conceptual bridge between the research in biology, pre-biotic chemistry, biomimetics, and the tools used in organic bioelectronics in terms of materials and devices. The goal is discussing how materials and devices of organic bioelectronics can be exploited and used at the interface with biology, but also how, and at what extent, they can be adapted to mimicking nature-inspired properties, herein including unconventional computing strategies. The idea is to provide new hints and solid hypotheses for designing niche experiments that could benefit from a proper interaction, even at a basic communicative level, between materials science and biotechnology. The finale long-term vision goal being the vision of collecting experimental data that may help to made a step forward toward the implementation of the transition from inanimate objects to animated beings. The mathematical model canonically considered in this work is the Navier-Stokes-Nernst-Planck (NPNS) Model which is often used to model a charged continuum system such as the organic electrochemical transistors.}




\maketitle

\section{Introduction}\label{sec1}

Understanding the potential of organic bioelectronics (OBE) in addressing fundamental questions in the fields of prebiotic chemistry and the origins of life opens many exciting avenues for interdisciplinary research. Can OBE serve as a complementary platform for exploring open problems in biomimicking, synthetic biology, and astrobiology? Furthermore, could the integration of biological materials with organic devices enable the creation of hybrid systems, where the active biological elements not just participate but also actively define the behavior of these devices? These questions form the underlying basis of this work, aiming to bridge the gap between bioelectronics and prebiotic chemistry through offering innovative strategies and insights. \newline

Organic bioelectronics was born as a natural evolution of molecular and organic electronics. It exploits the properties of novel materials such as polymers, small molecules, biopolymers and 2D materials, to develop devices which are then interfaced with biological systems. Devices developed thereof have been applied directly on the surface of the brain for recording EEG signals \cite{Liang2021, Khodagholy2013}, subsequently used for studying the response of cells cultures \cite{Lin2010,Romeo2015}, and applied directly over the skin for a long-term biopotential recording (ECG) \cite{Yang2023}. These results were made possible thanks to the design of dedicated organic devices and, as mentioned, by the properties of the underlying materials. One of the most important class of electronic devices for application in bioelectronics is given by the organic transistor based on the electrolyte-gating concept \cite{Lin2012, Tarabella2013}. Herein, we focus on the organic electrochemical transistor (OECT) because of its noticeably wide applications in biology \cite{Wang2021, Ohayon2023}  and potentiality in unconventional computing (UC). In particular, we summarise the basics referring to OECTs potential when applied at the interface with biological systems, in view of designing experiments of interest for the UC. \newline 

An Organic Electrochemical Transistor (OECT) operates through the interaction between ionic and electronic charge carriers within an organic semiconductor \cite{Ohayon2023}. The basic structure of an OECT includes an organic semiconductor channel, typically a conducting polymer like poly(3,4-ethylenedioxythiophene) doped with poly(styrene sulfonate) (PEDOT:PSS) or n-type conductors, \cite{Giovannitti2016} positioned between two electrodes called source and drain. The gate electrode, which is in contact with an electrolyte, is responsible for modulating the charge carriers in the channel. The gate voltage controls the extent of doping/de-doping of the organic conductor, thereby regulating the current flow between the source and drain electrodes. The modulation of the channel's conductivity depends on the movement of ions from the electrolyte and their interaction with the organic semiconductor. The OECT's operation is characterized by a mixed conduction, meaning both ionic and electronic charges determine the overall channel current. The responsiveness to ionic currents makes the OECT suitable for applications in biosensing, in particular in real-time monitoring of biosystem activity \cite{Butina2021, DAngelo2017}, where the detection of biological signals can be converted into measurable electrical signals. OECTs typically operate at low voltages, with compatibility to aqueous environments, and offer high transconductance/amplification in ion-to-electron transduction, making them effective for flexible, biocompatible electronics. Accordingly, the main strength of OECT are: 1. Low Operating Voltage (1 V), which makes them energy-efficient and compatible with the biointerfacing, as the operation in presence of biofluids is guaranteed at bias voltages below 1.23 V (the onset of water hydrolysis) \cite{DAngelo2019}; 2. High Transconductance due to the volumetric interaction (also known as ‘volumetric capacitance’), between ions and the channel, which allows an efficient current modulation \cite{Khodagholy2013_2}. 
In fact, PEDOT:PSS acts as a volumetric capacitor, hence interfacial properties upon interaction with an ionic reservoir are amplified by the polymer bulk; 3. Biocompatibility due to the use of organic semiconductors opens the way towards biointerfacing, hence making OECTs suitable for bioelectronics and biosensing applications. The combination between the PEDOT:PSS properties and device architectures (whereas, also true 3D  prototypes are possible due to 3D printing routes \cite{tarabella2019multifunctional}) allow to impart a  neuromorphic function to OECTs. Unlike electrolyte gated organic transistors exploiting the interfacial field effect to operate the channel current modulation, OECT's bulk operation involves strong materials \cite{WintherJensen2015} and geometry \cite{Ji2021} dependence, hence the device operation window, as well as both the dynamics and efficiency of recovery upon gate voltage switching off, can be easily tuned in a controllable way. 
\begin{figure}
  \centering
  \includegraphics[width=0.95\textwidth]{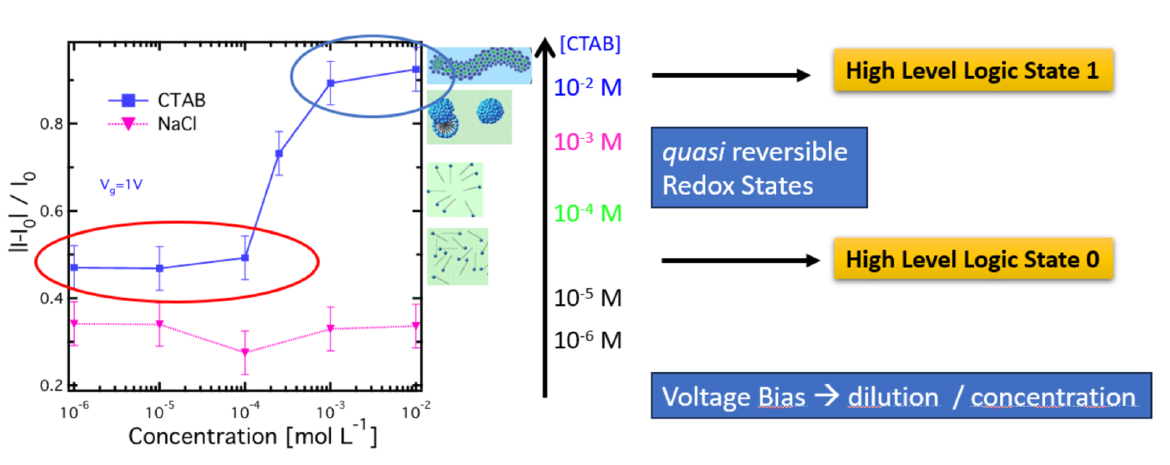}
 \caption{The OECT current modulation parameter as a function of surfactant concentration displays a transition from a low modulation region (off state, red circled region) to a region where a high modulation parameter if found (on state, blue circled region). } 
  \label{fig:fig1}
\end{figure}

\section{The nanogap OECT as a bridge toward applications in UC}\label{sec:sec2} 
The OECT architecture in terms of channel size/geometry that is typically explored in most literature studies is well suitable for applications with biological systems; nevertheless, it shows a low switching dynamics, falling in the range of seconds, as well as limited cycles of operations before detriment of performances. Therefore, literature standard OECTs are definitely unsuitable to be used as units in logic circuits, calculation and more in general, for UC applications. To overcome these issues, D'Angelo et al. developed a nanogap design of OECT \cite{DAngelo2019_2}. The nanogap-OECT shows a higher commutation speed that makes it more suitable for applications in logic gates. More, it is ideal for the interfacing of the active channel with single biological units with dimensions falling in the sub-micrometer scale. The channel scaling down to the to the nanometer scale allows enhancing performance by: (i) increasing the electric field strength across the channel, which influences the field-dependent mobility of charge carrier mobility in organics, and decreasing the drifting time of charge carriers from source to drain; (ii) reducing the gate voltage-assisted ionic diffusion dynamics within the polymer bulk, whose weight into the economy of the device response in time is critical in determining a diffusion-dependent dynamics \cite{DAngelo2019_2}. The faster device responses and greater sensitivity owed to the channel scaling promote OECTs as ideal platforms for applications such as biosensing, neural interfaces and bioelectronics. The nanogap design facilitates a better control over ion transport, resulting in more precise, reproducible and efficient signal conversion. An advanced vertical configuration known as vertical OECTs (vOECTs) also employs nanogap channels and leverages a stacked structure to reduce the channel length to sub-micron levels, often around 0.1 $\si{\micro\meter}$ \cite{Huang2023}. This design improves switching speeds and transconductance, while providing greater stability and scalability compared to traditional planar OECTs. Here, we aim to expand the conceptual framework by suggesting that OECTs traditionally used to study biological systems, can also be applied to investigate pre-biotic systems. This approach would allow researchers to explore the behaviour of fundamental biological components and their interactions. OECTs could be utilized to monitor how these pre-biotic molecules respond in the context of collective interactions demonstrated in literature \cite{Mougkogiannis2023}, track their potential evolutionary changes over time, and evaluate their reactions to electrical stimuli. By detecting these interactions at an early stage, OECTs may provide insights into the underlying mechanisms of molecular evolution, offering a deeper understanding of how fundamental biological components may have functioned or evolved in primordial environments. The OECT operation scheme under the real time or continuous operation, usually implemented in typical Lab-on-Chip microfluidic/OECT platforms, indeed represents the key point for such a class of devices, due to the fast doping/de-doping dynamics of PEDOT:PSS, compared to that of other electroactive polymers, such as polyanilines \cite{Demin2014}. It is evident that all the features discussed so far are intimately connected to the properties of organic semiconductor materials, which are biocompatible, biodegradable and, most importantly, exhibit a mixed ionic/electronic conduction.  The reference materials, such as the above-mentioned PEDOT:PSS, n-type polymers based on the  naphthalene diimide monomer and PANI, but also polypyrroles (PPy), polythiophenes (e.g. P3HT) and other electroactive conductors, are suitable for achieving all of this, and their chemical, physicochemical, and biochemical modifications may easily enhance their potential. At the end, organic semiconductors can play a central role due to their unique and versatile properties.

\begin{figure}
  \centering
  \includegraphics[width=0.95\textwidth]{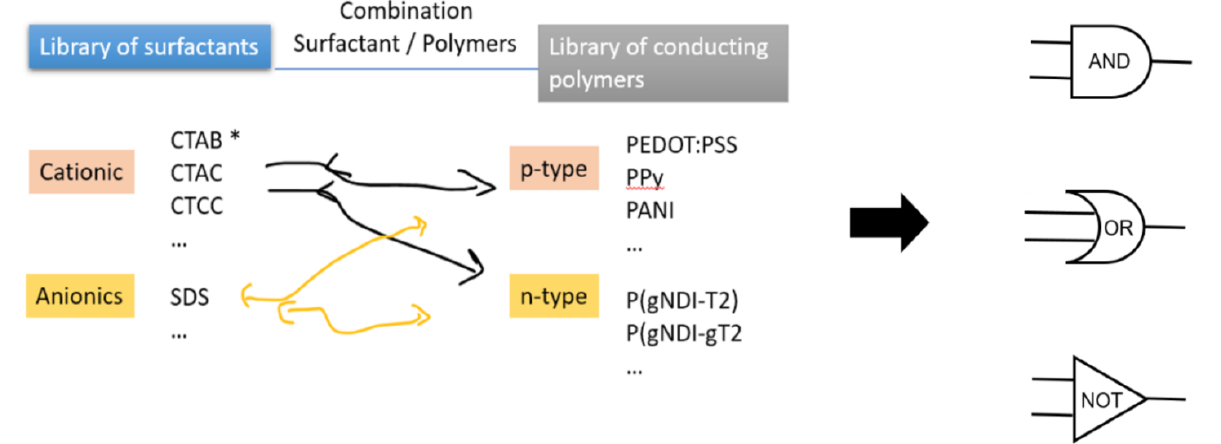}
 \caption{Library of surfactants and organic semiconductors to implement logic gates and related combinations.} 
  \label{fig:fig2}
\end{figure}

\section{Living hybrid devices}\label{sec3}
Living hybrid devices exploit biological structures as their active element, i.e., the device itself is made of a biological part realizing a bio-interface capable to actively contribute to the device operation as a sensor and/or as a neuromorphic device. Literature offers some examples of living hybrid devices where a living cell has been interfaced with an organic conductor and acts as an active element for the device operation. \newline 

A first example is given by a hybrid bio-organic electrochemical transistor, developed by integrating the organic semiconductor PEDOT:PSS with an eukaryotic macroscopic cell, i.e. the Physarum Polycephalum slime mould (PPC) \cite{Tarabella2015}. This innovative system, developed in the margins of a funded FP7 EU project, named PHY-CHIP, exhibits exceptional performance in so far as it can operate both as a transistor in a traditional three-terminal configuration and as a memristive device in a two-terminal setup. This dual functionality is highly significant, as the transistor mode enables the system to act as a highly sensitive biosensor, capable of directly monitoring biochemical processes within the living cell. Meanwhile, in memristive mode, it stands out as one of the first bio-hybrid electronic devices demonstrating memory capabilities. The unique combination of memory and sensing in a single device opens new possibilities for designing and manufacturing unconventional computing platforms. A full electrical characterization of the system has been reported in experiments conducted using different gate electrode materials, i.e., silver (Ag), gold (Au), and platinum (Pt)—which are known to exhibit distinct behaviours in organic electrochemical transistors. The performed testing measurements reveal that the proposed system holds significant promise for advanced sensing applications. This hybrid system can be classified as a Bio-Organic Sensing/Memristive Device (BOSMD), where its dual nature merges sensing and memory functions, potentially unlocking new opportunities in bioelectronics and paving the way for unexplored applications at the frontier between both the biosensing and neuromorphic device-related fields.
Another significant case of hybrid devices having a configuration similar to that of PANI-based memristors in a three-terminals electrodes configuration (i.e. a silent gate electrode in an OECT architecture), again using the PPC slime mould as a living bio-electrolyte. This hybrid organic/living prototype displays a synaptic-like behaviour \cite{Romeo2015_2}. A conductivity switch can be induced as a consequence of an over-oxidation process regulated by the application of a channel voltage . The over-oxidation is due to the ionic flux taking place at the PPC/polymer interface. This behaviour endows a current-depending memory effect to the device mimicking the long term synaptic plasticity of neurons. \newline

The possible implementation of the above described devices in hardware networks may offer the opportunity for merging recognition capabilities and data classification by electronic devices, hence at the hardware level.  These examples of hybrid living being devices have been by fact recognized as pioneering literature examples of neuromorphic transistors based on the OECT concept \cite{Gkoupidenis2015}, and they also confirm the high versatility of OECTs in terms of multifunctionality integrated in one single device with a transistor-like operation, at the basis of digital electronics and computational network implementation. The materials used in Organic electronics are carbon-based polymeric semiconductors, the ideal candidates for implementing an interfacing with living systems due to the mixed conduction, which mimics the basic features of biosystems’ activity, also integrating at the same time high flexibility and chemical compatibility with the biologic world. Within the biomedical research field, such materials are actively employed to study the basic mechanism and monitor the evolution of biological processes, accordingly representing tools suited to enhance the understanding of the electrophysiological properties of living systems. Bioelectronics can be contextually used for addressing specific and poorly understood mechanisms at the basis of life, as well as at understanding how to exploit such properties of basic and simplest life blocks in order to determine tools suitable for concretely implementing an unconventional computation. The assembly of materials in a first protocell to reproduce in a laboratory environment a sort of transition from non-living to living matter, following a mechanism which is argued to be at the basis of the origin of life, is a challenging discussion. Currently, the problem is still remaining at the borderline with philosophical concepts. The main problem is that it is difficult to design and perform experiments able to deliver a set of reliable data from basic structures of polypeptides spontaneously self-assembled in a globular-like configuration (proteinoids). This is essentially owed to the scarce attention given by researchers to the molecular evolution of random or disordered globular proteins, a mechanism believed to be at the basis of the origin of life, with respect to complex biomolecules as nucleic acids, which are by fact at the basis of all the living beings. \newline

Conducting polymers show some of the main characteristics of living systems that contribute to formulating a paradigm involving living beings and bioinspired materials engineering; in particular, they show self-repair and self-reproduction capabilities, together with the adaptability in response to the environmental stimuli. For example, under specific conditions the PEDOT:PSS polymer also has a sort of self-repair property that mimics a fundamental property of living beings, i.e., the self-repairing capability of tissues (e.g., wound healing) \cite{Zhang2017}. This outstanding property reinforces the field of action allowed by the PEDOT:PSS polymer, integrating the as-discussed properties of biocompatibility, useful for implementing the interfacing with biological systems,  and efficient mixed ionic-electronic conduction. This is just an example of a possible connection between the capacity of living systems (self-healing) and the morphological property offered by a material in the organic bioelectronics portfolio, in view of the implementation of hardware systems that may mimic the functions of organs. 

\section{Unconventional Computing with biological materials}\label{sec4}

Unconventional computing using biological materials is an emerging field that explores the potential of living systems and biological molecules to perform computational tasks. Unlike traditional silicon-based computers, biological computing harnesses the inherent complexity, adaptability, and parallelism of biological entities like DNA, proteins, and even entire cells. Applications of this approach include bio-sensing for environmental monitoring, smart drug delivery systems that respond to specific molecular signals, and biocomputing devices that can be directly and easily interfaced with human tissues for advanced medical diagnostics. By leveraging the unique properties of biological materials, this field aims to create highly efficient, sustainable, and innovative computational solutions that could revolutionize technology and healthcare. In the following, we propose an experimental design consisting of a coupling between complex biological molecules and semiconducting polymers to perform basic computational tasks.

\subsection{Synthetic biomolecules: micelles}

Surfactants are compounds that lower the surface tension between two substances, such as liquids and solids, making them essential in a wide range of applications from cleaning agents to pharmaceuticals. The critical micellar concentration (CMC), defined as the concentration at which surfactant molecules begin to aggregate into micelles in solution and above which also different aggregates in terms of morpho-structures may be found, is a key parameter in surfactant chemistry. An OECT can serve as an effective sensor for detecting the CMC of surfactants. By using the OECT with a surfactant-sensitive material as electrolyte, the electrochemical properties of its active polymeric channel are tuned to respond to changes induced in the surfactant concentration. As the surfactant concentration approaches the CMC, interactions between the surfactant molecules and the sensor material alter the OECT’s electrical characteristics, enabling precise and real-time monitoring of the CMC. An OECT made of PEDOT:PSS can detect the CMC of a cationic surfactant as CTAB \cite{Tarabella2012}. \newline

Fig.\ref{fig:fig1} shows the electrical response of OECT after exposure to the cationic surfactant CTAB to a simple salt, such as an aqueous solution of NaCl. The OECT response is recorded as the current flowing along the PEDOT:PSS channel. Above the CMC, the micelles of CTAB salt can efficiently de-dope the PEDOT:PSS such that the current modulation increases. The de-doping effect can be ascribed to the surface charge of micelles as measured by Dynamic Light Scattering.  \newline

To implement an unconventional logic state, by exploiting the state of aggregation of a surfactant, we may consider the 2-level of current modulation related to the dissociated vs. aggregated state of the surfactant, as 2-logic states of high (level 1) vs. low (level 0) current values representing the binary logic states units. By exploiting the ability of a PEDOT:PSS-based OECT to detect the CMC of a surfactant, we aim at implementing a reversible 2 logic states switching by exploiting soft matter through a simple water-dilution. Under this unconventional computing scheme, the standard voltage bias on/off is replaced by a water dilution/concentration. \newline

Computational logic can be made more sophisticated in steps of increasing complexity: (1) by using surfactants at different CMC; (2) by coupling cationic and anionic surfactants, (3) by implementing OECT with n and p-type polymeric channels. For both surfactant and polymers, we can envisage a library of materials, the combination of which could allow to design complex computation tasks and building logic hardware (Fig. \ref{fig:fig2}).\newline

\subsection{Natural biomolecules: pigments}
\begin{figure}
  \centering
  \includegraphics[width=0.95\textwidth]{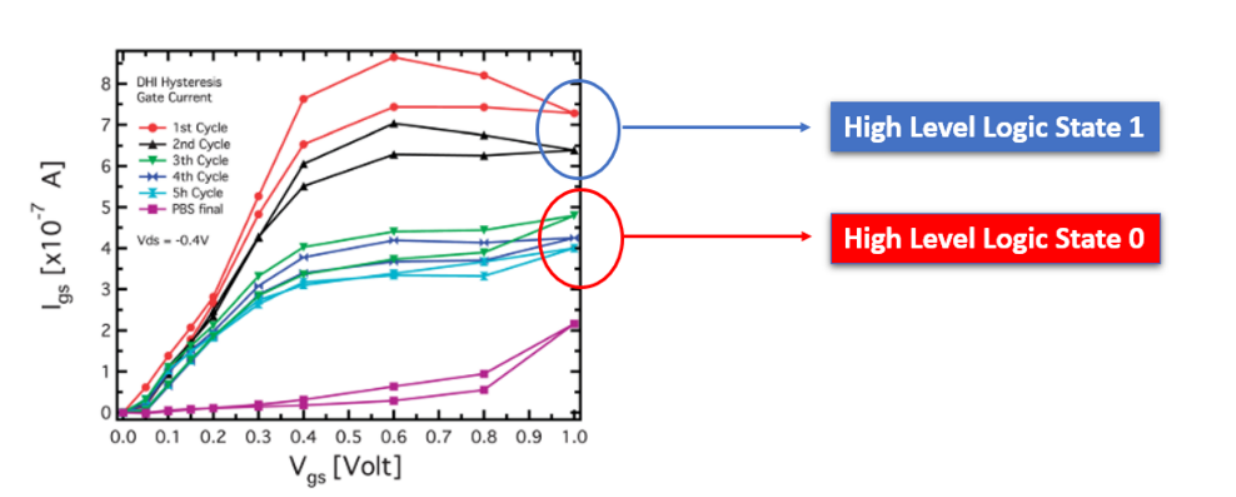}
 \caption{Hysteresis loop cycles of the gate current acquired for DHI.} 
  \label{fig:fig3}
\end{figure}

The above example shows how to implement a device for UC scope simply using a synthetic biomolecule. However, based on the use of appropriate biological materials, it is also possible to implement reversible and irreversible logic states by an OECT. Pigments of melanins, such as eumelanin, and indoles involved into the melainins’ biosynthesis, such as the  5,6-dihydroxyindole (DHI), have shown the ability to de-dope PEDOT:PSS channels in OECT \cite{Tarabella2013_2}. Upon cyclying the gate bias within a restricted voltage window, it is possible to discriminate a high level of current modulation, associated to the logic state 1, with respect to a low level achieved after thee cycles of hysteretic modulation of the gate current. Herein, as for the previous case, the combinatorial coupling of different types of materials (polymers n and p) would allow the opportunity to build an array of logic gates (Fig. \ref{fig:fig3}). Furthermore, the logic gate operation is implemented by the gate circuit, leaving the electronic circuit schematizing the conduction channel available for implementing an integration in complex logic gates and artificial neuronal networks. \newline

\section{NPNS model for OECT systems: mathematical and analog computing methods}\label{sec5}

The question, posed in Sec. \ref{sec1}, also invites us to identify a canonical model that can seamlessly collate the study of pre-biotic chemistry (which is post planetary-accretion) and the study of life (as it appeared on Earth). One model that we consider in our discussion is the Nernst-Planck-Navier-Stokes (NPNS) model. NPNS model is used to conservatively define a system consisting of: (1) multiple species of ions with respective charges $z_{i}$ (and corresponding signs/valencies), concentrations $c_{i}$, and diffusivities $D_{i}$, such that the bulk of this system contains, (2) an incompressible fluid with constant density $\rho$, viscosity $\mu$, and velocity $u$, and lastly, (3) an electric field generated because of local charge $e$ and applied voltage $\phi$. In Constantin-Ignatova-Noah Lee (CIN) \cite{Constantin2021}, the global smooth solutions for such a system were investigated. Under certain assumptions, the authors were able to establish global regularity of the global smooth solutions for arbitrary smooth data in the bounded domains. We now pause to explain these mathematical terms, one-by-one, in the bioelectronic and OECT context for our readers. To start with, the NPNS model is given by a system of equations, which combines the dynamics of migration of electrolytes $(c_{i},D_{i}, z_{i})$ with the flow of charge carriers $(e,\phi)$ and the continuum flow of organic material $(\rho,u,\mu)$, across the organic semiconductor channel. The channel is geometrically defined as a specific type of domain $\Omega$. For a given $\Omega$, boundary conditions for $c_{i}$,$\phi$  are given by inhomogeneous Dirichlet conditions. The result obtained in \cite{Constantin2021}, is that the semilinear parabolic systems of equations, which model the flow of a “charged material” continuum inside the channel whilst obeying boundary conditions and initial data which are smoothly describable, tends to have a well-defined smooth solution inside the entire channel, given that a certain necessary-and-sufficient condition is met. Once such solutions are obtained, it is possible to describe them computationally, thanks to the Church-Turing thesis, should there be any need for it. The need, though, becomes necessity once the equations get complicated enough (though still belonging to the same class as the NPNS system) and computational power required to obtain the solution becomes a bottleneck. \newline 

This takes us to the next series of discussions on standard Turing-complete and non-standard (analog unconventional) computational methods to model OECT systems. As alluded before, smooth regular solutions to the NPNS system of equations were derived by CIN. Smooth functions are, by definition, infinitely-differentiable functions, whose existence fully ensures that our model is physically reasonable and reliable. Then the next thing to worry about is the accuracy of such models. Do the models output the value (the solutions of the system of equations) that is significantly close to the experimental observations? However, an astute reader will likely intervene here and ask if the NPNS system of equations models all kinds of OECTs observed in practice, particularly the ones which do not obey the assumptions adopted at the first place. Such complex kinds of OECTs, could, for example, have the following properties: (i) more than two charged species, (ii) varying diffusivities of charged species, (iii) non-smooth boundary data, which is enough to violate the assumptions used in CIN. Another key assumption that can hold back other morally equivalent results as observed by CIN is the bounded-ness of the domain (where the charged continuum material lives). Had this assumption been lifted off, such as in an infinitely spatial domain, then the (in)famous global regularity questions pertaining the Navier-Stokes equations would creep into the discussion, and because the problem of global regularity of Navier-Stokes equations remains open \cite{Tao2016}, the general problem of solving NPNS system would remain morally open. \newline 

Before further discussing the computationally programmable nature of the solutions to OECT systems modelled by the NPNS set of equations, it is important to see what is the status quo of solutions to the physical systems modelled by Navier-Stokes (NS) set of equations, and then expanding further the insight, from there on to a complete NPNS system (for a bounded domain where the charged continuum material lives). After the proof of blowup for artificial NS equations, which were interpreted as a quadratic circuit of logic gates (pump, amplifier, and rotor gates), Tao \cite{Tao2016} suggested further investigating the computing capabilities of a continuum fluid, using logic gates, which perhaps could have a cellular automaton, von Neumann machine, or quantum computation like programmable features. This idea was investigated by Miranda-Cardona-Peralta Salas-Presas \cite{Cardona2024} where the authors proved that an inviscid class of NS equations, on a Riemannian 3D sphere, are Turing complete and programmable. The programmability of much broader class of equations than the inviscid NS equations were thoroughly surveyed in a recent article by Cardona-Miranda-Peralta-Salas \cite{Cardona2024}. It remains an open avenue to expand these computability notions to equations which are coupled with Navier-Stokes, such as the Nernst-Planck-Navier-Stokes equations, particularly for the cases where global regularity is not well-known \cite{Constantin2021}. Now, let’s take a step back, and ask ourselves: where does an inherent computability in the “charged continuum” systems take us? One likely direction is towards the well matured and ripened field of super-Turing systems, such as: analog computers, transformers, LLMs, and within the analog computing systems, numerous paradigms such as: neuromorphic computing systems, reservoir computing, and analog neural networks lie. These analog systems are typically classified under the umbrella of unconventional computing systems, which are not restricted to physical systems, and thus spans the vast array of biological, biochemical, and chemical systems. To give an example, Sharma-Marcucci \cite{Sharma2022} explored the avenues of computing for uncharged continuum systems, by drawing attention specifically on Liquid State Machines (LSM), Echo State Networks (ESNs), Extreme Learning Machines (ELMs), and in general, Neuromorphic Computing (NMC). We leave as an exercise for the interested reader to investigate the avenues of computing for charged continuum (OECT) systems in a similar fashion as previously done for uncharged continuum systems.

\section{Mathematical results on the NPNS modelling: a summary}
This section outline some of the mathematical results related to the NPNS modelling. Global existence and stability for the no-flux, blocking boundary conditions has been studied previously in \cite{biler2000long}, \cite{choi1995multi}, \cite{gajewski1996reaction}, for 2D, 3D (small data) cases, or in a weak sense; uncoupled with fluids. Global existence for weak solutions for equations coupled with fluids, for homogeneous Neumann boundary conditions on the potential was previously studied in \cite{schmuck2009analysis}. For homogeneous Dirichlet b.c. on the potential, global existence of weak solution in 2D for large initial data and in 3D for small initial data has been analysed previously in \cite{ryham2009existence}.  The case of Robbin b.c., with global regularity problem in the 2D case is studied in \cite{bothe2014global}. Global existence of smooth solutions for blocking b.c. and for uniform selective special stable Dirichlet b.c. in 2D space dim was obtained in \cite{constantin2019nernst}. When b.c. is general selective, numerical studies (\cite{rubinstein2000electro}, \cite{zaltzman2007electro}) and experiments (\cite{rubinstein2008direct}) show instability. It is useful to draw a visual diagram of the results outlined above, more aligned in spirit with the suggestion outlined in this \href{https://terrytao.wordpress.com/2025/01/28/new-exponent-pairs-zero-density-estimates-and-zero-additive-energy-estimates-a-systematic-approach/}{blogpost} for number-theoretic results. 

    \section{Conclusions}\label{sec6}

Materials science is currently making a massive use of synthetic and natural polymers to respond to the request for introducing sustainable raw materials and manufacturing methods for the development of electronic devices and related application in bioelectronics. A niche research is however devoted to the use of natural or synthetic biomolecules, living beings and basic constituents of life, of interest for pre-biotic chemistry, to implement (bio)electronic devices that are able to perform a basic transistor function coupled to the peculiar capability at operating as biosensors, due to materials biocompatibility and allowed operation in fluid environment. Biomolecules like surfactants and pigments, spherical proteinoids and eukaryotic cells may play an active role in transistor operation that, in case of the specific OECT platform, has demonstrated to be effective in implementing the study of biosystems properties under a continuous mode of operation. In this commentary, we have explored how these multifunctional devices can exploit such properties to fulfil the main requirement of a computation based on unconventional mechanisms. These aspects may be accomplished because of the operational multifunctionality by organic and hybrid organic-living being devices. In this respect, the discussion among scientists may help to find a way to accomplish specific needs of a research field like the UC upon exploiting the multifunctionality of organic devices on different levels. In this perspective, based on our own experience, we have discussed through simple examples how a specific device like the versatile OECT, while implementing a transistor function, may be a tool for studying and exploiting biosystems in concrete computational tasks. This specific goal has been also extended to some perspectives that may arise from the connection between OECTs and pre-biotic chemistry, whereas some studies related to the occurrence of electrical spiking from collective oscillation by periodic spherical proteinoids have been already presented in literature, and indicated as promising for UC. However, approaches in terms of modeling that are indeed required to well adapt the OECT operation with findings of pre-biotic chemistry, have been discussed too.

\begin{appendices}

\section{}\label{secA1}

A brief outline of proof of the global regularity of NPNS system of equations, for bounded connected domain $\Omega$. Let us state the result observed in \cite{Constantin2021}. \newline

\textbf{Theorem 1.} Let $c_i(0) - \Gamma_i \in H^1_0(\Omega)$, and $u(0) \in V$. There exists $T_0$ depending on $\| u_0 \|_V$ and $\| c_i(0) \|_{H^1(\Omega)}$, the boundary conditions $\gamma_i, W$, and the parameters of the problem $(\mu, K, D_i, \epsilon, z_i)$, so that the Navier--Stokes--Nernst--Planck system with appropriate boundary conditions for $c_i$ and $\phi$ has a unique strong solution $(c_i, \phi, u)$ on the interval $[0, T_0]$.

\textbf{Scheme of the Proof.} The proof of Theorem 1 establishes the existence of a unique strong solution for the Nernst--Planck--Navier--Stokes system over a short time interval by using a sequence of steps to control the solution's behaviour. First, the concentration $c_i$ is decomposed into a known component $\Gamma_i$ and an unknown remainder $q_i$, simplifying the equations. The reformulated equations, now focused on $q_i$ and other system variables, are solved under homogeneous Dirichlet boundary conditions, which help in managing the estimates. The proof then derives \textit{a priori} estimates for each term in the equation to show that the solution's energy (in terms of $q_i$, $u$, and related terms) remains bounded over time, preventing any ``blow-up'' in values. By combining these estimates, it demonstrates that all quantities remain under control on a small interval $[0, T_0]$, which depends on initial conditions and system parameters. This control ensures the existence and uniqueness of a strong solution over the interval, thus proving the statement above.

\end{appendices}


\bibliography{sn-bibliography}

\end{document}